\documentclass[aps,pre,twocolumn,floatfix,showpacs,showkeys]{revtex4}
\usepackage{amsmath,amsfonts,amssymb,mathrsfs,graphicx}
\begin{document}
\title{Chaos and Annealing in Social Networks}

\author{Fariel Shafee}
\affiliation{ Department of Physics\\ Princeton University\\
Princeton, NJ 08540\\ USA.} \email{fshafee@princeton.edu }

\begin {abstract}
In this work we compare social clusters with  spin clusters and
compare different properties.   We also try to compare phase
changes in market and labor stratification with phase changes of
spin clusters.  Then we compare the requisites for redrawing the
boundaries of social clusters with respect to energy minimization
and efficiency. We finally do a simulation experiment and show
that by choosing suitable link matrices for agents and attributes
of the same and of different agents it is possible to have at the
same time behavior similar to chaos or punctuated equilibrium in
some attributes or fairly regular oscillations of preferences for
other attributes, using greatest utility or efficiency as a
criterion for change in conflicting social networks with different
agents having different preferences with respect to the
attributes in the agent himself or with similar attributes in
other agents.

\end{abstract}

 \vspace*{1cm}

\maketitle

\section{Introduction}

In economics, every agent behaves rationally to maximize his or
her utility curves.  However, in an imperfect world, not all the
people we know look rational.  Phenomena like war and spite look
very contrary to what a rational person would do to maximize his
or her material needs.

The reason human interactions are possible, and it is also
possible to have a coherent system of trade and negotiation is
explained by the idea that a common set of needs and understanding
are shared by the majority of people.  Classical economics deals
with this set of needs and formulates equations for every possible
deal.  However, it fails to account for the so called irrational
human behaviors.  If a set of scientific data observed by most and
possible to be experimentally verified by the majority did not
exist, social networks would perish.  However, together with the
existence of a network with a common set of knowledge, exist
phenomena like revolutions, hatred, riots, and many basic
instincts not explainable by a concrete set of knowledge. Attempts
have been made to justify irrational behaviors as weak,
inconsistent or culturally inferior and evolution-wise unfit.
However, in this paper, we try to understand the fine structure of
human perceptions and needs to model possible chaotic and
annealing behaviors of social lattices from a spin glass point of
view without any a priori bias towards a certain logic system
successful in adapting with nature at a certain time point.

Previously, we have tried explaining the non quantifiable human
needs \cite{FS1} and the variations in utility curves by taking
G\"odel's incompleteness theorem \cite{GO1} into account. In
discrete math only the states 0 and 1 exist for switching. In the
real world, of course, there can be many states in between
corresponding to a continuum. However, one can also consider only
extreme conflicting cases that are opposites in a simplified
picture. We have argued that although strategies to maximize needs
might be rational, the needs themselves might not be a "common
rational" set of constants. As any logic system is based on a set of
axioms that are by themselves not provable within the logic system
itself, and the axioms or sets of information are acquired by each
agent individually by interacting with its environment and other
agents, not all agents necessarily need to share the same set of
axioms or needs that they tend to optimize. Some needs, however,
might be very common and "material" whereas some other needs are
quite hard to quantify.  We try to explain the branchings of these
spin clusters to chaos and annealing by taking this unprovability
theorem into account.  We argue that although majority of the people
must agree on a majority of utilities and perceptions, fine
structures in variables that define agents and their perceptions or
utilities connected with other variables in a lattice may cause
chaotic behavior or annealing when placed in an interacting lattice.

In our simplified model, we describe each agent as an n tuple of
qualities and needs.  We call each of these entries in an "array"
possessed by the agent as a variable.   We start with a set of
basic quantifiable needs that an agent tends to maximize in this
paper.  We also define a set of variables that an agent possesses
that he/she can use in order to maximize the needs.  The second
set will include skills that the agent can trade and calculating
efficiency to interpret the other agents' needs as subsets.  The
third set will define a set of easily discernible variables that
other agents can use to mark or tag an agent.  We can then
describe each agent as a multivariable spin state, and investigate
how it interacts in a social cluster and with nature.

\section{The Definition of "self"}
We define an agent's "self" as an array of variables possessed by
the agent.  In our model, the idea of self, instead of being an
object contained within a body, will be rather a slowly weakening
concept of the self similar variables.  We use the term slowly
weakening because the idea of "self" decreases as another array
contains fewer variables in common and also the strength of the
"bonds" as a function of physical or need wise distance.   By
need-wise distance we mean the placement of a variable in a
position where it can be influenced by the agent or from where it
can influence the agent.  This distance is not necessarily solely
a geographic distance.  As a result, our model of "self" is a
fractal like model with a basic array of variables forming the
shortest unit.   As we include more and more self similar
variables from other agents placed in a cluster but also containing dissimilar variables,
the idea of self weakens.

 The "existence" axiom  is related to a concept of "self".
 But in this paper, we propose making the definition fuzzy.
 Networks arise because people  associate with others who are similar
 to themselves in some respect  \cite{LM1}.
 In our model, each agent can be defined as finitely or infinitely
 many variables in an n-tuple or an array.  An agent will identify
 the array with the definition of self.  However, the variables are
 connected to the concept of "self" with certain weight factors which
 maybe modifiable.  An agent chooses to maximize the utility curves
 for him"self".   He may choose to perpetuate different variables
 related to "self" with more energy invested.  This preference,
 again, is complex, and may not be definable by simple functions.
 However, the other agents in the cluster also possess an array of
 variables and may or may not have the same value as the first agent
 in each of the variable slot.  We can say that an agent will
 "identify" with other agents who possess similar values for
 the variables.  This extended fuzzy idea of "self" may create
 some loosely or strongly defined clusters where the agents
 relate with self-similar structures in specific variables. The affinity
 with agents may also be a function of time, as which variables an agent
 would tend to find more important may be a dynamic function of time.
 When an agent invests in perpetuating "self", we thus add a correction
 term to the self that includes other agents possessing similar variables.
 However, granted that no two agents are similar in every possible
 variable, any investment in perpetuating other agents with similar
 variables also comes with the cost of perpetuating some dissimilar variables.

These affinity clusters may be modeled as follows: We can describe
an agent as an array of variables.  The variables may have discrete
0 or 1 values or continuous values within a range. Within the social
lattice, we can say that each of these spin type variables is
coupled to the similar variable slot of other agents. The coupling
will depend on a coupling strength, which might be a function of
time, and also on the values of the variables in the two agents.
What is interesting here is that the agents may or may not have
knowledge of the proper values for the variables.  So each variable
in an agent has an actual value and also a value perceived by the
other agent/s.  The individual utility curve can get distorted
because of this "affinity factor" as the corrected utility would be
the utility of the individual corrected by a weighted utility of the
affinity group.  The affinity factor will reflect an agent's
perception of other agents' variable values.  A miscalculation or
bluffing on other agents' parts will cause the agent to invest
energy in perpetuating dissimilar or contradictory variables.
However, each member within a known affinity group will have the
individual affinity and the corrected group affinity both playing a
role in the utility curve.  Moreover, the group utility is a factor
that is shared by all members of the group, and each individual will
tend to utilize the other agent's affinity points to gain an
increase in utility at no or little expense of ones own cost of
entropy.  As a result, it might be profitable to "net utility" to
invest in increasing other agents' affinity utility factor. Again,
investments made in maximizing group affinity utilities from others
will result in an expenditure within a closed subdomain.  If the
total amount of investment, or energy to be spent is kept constant,
an increased investment within the subdomain will reduce investments
in games with other subdomains and with nature.  As a result, agents
in a subdomain playing against each other with a large weight factor
to increase each others group affinity factor will have less to
invest in games outside the subdomain. Again, bluffing about ones
variables might be a strategy an agent uses to gain group affinity
points from agents with dissimilar variables with no expenditure of
group affinity from the agent's own side. We can check out by
simulation cases where agents who are indeed sharing affinity
variables are put together in a cluster, and clusters where some
concealing agents are mixed.

The other interesting property of this notion of "self" is that
the variables in the unit array are dynamic.  As a result, the
idea of self will change with time as external variables change.
However, as the basic array gets modified, so does a person's
perception of "self" and the affinity clusters that appear as
weaker versions of "self" also redraw their boundaries.

\section{VARIABLES AND THEIR EVOLUTION}
\subsection{Knowledge, Preferences}

Knowledge is a set of information that an agent obtains by
interacting with nature and other agents.  Knowledge is obtained
by symmetry breaking and provides an agent with a rule or a
measurement that is in a collapsed state and can be used in the
future to play against nature \cite{CO1}.  A knowledge or piece of
information can be used to satisfy the agent's preferences of what
should be done in the future. Although knowledge can be shared
among agents, so that agents can decide on a possible set of
largely overlapping common knowledge, the preferences about how
these knowledge pieces should be used to interact with nature is
not necessarily a common set of  knowns.  Preferences or
utilities, may vary in importance or weight from agent to agent.
Some of the utilities are fixed, and some are not.  Pieces of
knowledge are combined with resources that can be explained as
specific sites of interactions with nature can be used to change
nature.  Given a finite amount of resources, with many
possibilities of investment in future games, the "choice" of the
optimum investment remains unfixed.  Agents use the set of
knowledge they possess to interact with nature using the resources
to change nature to accommodate their preferences.  In appendix A,
we discuss a theoretical model for a group of agents possessing a
largely overlapping knowledge and preferences.

\subsection{Skill and Aptitude Variables}

The interactions of the agents with nature in order to modify
nature depend on both skill and aptitude.  An agent can modify his
ability to interact with nature by repeatedly interacting with
nature.  It is a mechanism of adaptation by which the agent learns
to optimize his efforts to interact in a certain way.  However,
the innate ability of an agent to interact in a certain way or
perform a certain deed can be contributed as talent or aptitude.
An agent with no aptitude will need to invest a higher amount of
time and energy to master a certain task.  We can model aptitude
as follows: We can assume that each agent is born with a group of
preference curves that are slightly different from other agents'
preference curves.  The curves can be modified by interacting with
nature, and can be brought to roughly resemble another person's
preference curve distribution.  However, adaptation and learning
will create an exact match with another distribution very rarely
with the increase in number of variables and steps in learning
process involved with a certain distribution.  A certain
distribution, gain, can be optimized to interact with nature in
specific ways and produce specific results.  A certain agent's
aptitude for a certain job may also be interrelated with his or
her aptitude for other specific jobs as performing a certain group
task optimally may require sharing one or more preference
variables.  However, these fine tunings are very rarely going to
match an exact optimized result need, and can be taken as a random
function of chance over all agents.

\subsection{Self and Mutability with respect to Fitness}

An agent will try to perpetuate him"self" and in order to do so,
he will play against nature and against other agents.  We tried
formulating this game in our first paper.  However, while playing
against nature, a certain set of variables will be fitter at a
certain time space point than other.  An agent will try to
dynamically change the idea of "self" to make himself fittest to
perpetuate.  On other words, an agent will either change variables
to be placed at an optimal condition with respect to nature, or
change nature to fit himself best.  As a second strategy, an agent
will try to include other agents with optimal variables in the
cluster of "extended self".  This can be done so by mating with an
agent with the coveted variable to include the variable in the
agent's own array.

\subsection{Optimized variables and Bluffing}
The agents use their set of variables to play against nature and
also against the other agents competing for a finite resource
offered by nature.  So although an agent may be interested in
utilizing another agent's more optimized variable, they are also
playing against each other to maximize the perpetuation of their own
variables.   Aa certain time point, an agent may not have a set of
variables that are all optimal with respect to nature. Agents can at
that point form a network where more optimized variables of one
agent are used by another agent in return for the other more
optimized variables of the second agent.  However, the possession of
more optimized variables puts an agent in a position with higher
bargaining power.  The game at this point can very easily be modeled
in the same fashion as several agents with cards with higher or
lower values.  An agent may show the value of the card before
placing a bid in the game or bluff.  In a pioneering theory on game
theory \cite{VO1} the authors argue how bluffing is an essential
strategy of any such game.  Similarly, in the game of social
lattice, equilibrium points are achieved by bluffing and trying to
interpret the correct optimizations of the variables possessed by
other agents.  The inclusion of an optimized variable in an agent's
array by mating with an agent possessing that variable, also
includes variables previously not in that array and not optimized
with respect to nature in the array of the agent. This new array
will for a new definition of self and will thus isolate the agent
from its older cluster defining self if the newly introduced
variables are dissimilar enough with the first set of variables.  As
a result, an agent not wishing to break his former "self" cluster
might as well find strategies to make use of the other agent's
optimized variable without making the second agent part of him
"self".

\subsection{Correlated and Uncorrelated Variables and Their
Interpretations}

Variables may be obtained by genetic inheritance or by
interactions with nature and with other agents.  Again, the
variables an agent possesses may or may not be correlated.
Moreover, they may appear to be correlated if they are placed in a
certain environment for a certain time period and then become
uncorrelated when the agent is removed from the environment.  For
example, if placed in a certain environment that requires
optimization of two variables for survival, an agent may develop
specializations in two factors.  However, if the agent is removed
from the environment, it is possible for one of the variables to
become randomized while the other remains fixed.  Again, some
variables may be dominant and others recessive.  As a result,
mating among specialized agents and non-specialized agents may
produce off-springs that are specialized in one variable but not
specialized in the other.

\subsection{Interpretation of Variables}

Some variables may be easy to understand whereas others may
require a heavier investment of time and energy for
interpretation.  We can say that some of the variables are easily
visible but others are not.  However, if two variables are
"assumed" to be correlated, interpreting the more easily
visible one saves time and energy that otherwise would have been
spent in interpreting the second one. This process may be efficient
as long as the two variables are in reality very closely
correlated, and the risk of the variables becoming uncorrelated is
very small.

\subsection{Difference with a Parochial Cluster}

Networks arise partly because agents choose to associate with
others who are similar to themselves in some great respect
\cite{LM1}. In Persistent Parochialism: Trust and Exclusion in
Ethnic Networks \cite{BO1}, the viability of ethnic clusters and
parochialism is described in detail.  We propose a model similar
in some aspects with the parochial model here, but also in terms
of the spin glass model.  Variables that are easy to fake and
variables that are difficult to fake are also taken into account.
However, the main dissimilarity between the Bowles model and our
model is that instead of taking parochial networks as networks
where the fixed variables act as markers for shared beliefs and
information only, we define an agent as an array of variables
where all the entries add up to the definition of "self" and the
agent finds strong or weak affinity with other agents depending on
the weight of the variable in the agent's definition of "self". As
a result, the agents do not form a network based on more easily
identifiable ethnic qualities because of the expectation of shared
information to enhance cooperation only. The agents will form
positive, negative or neutral bonds with all other agents
depending on the value of the variable at each variable slot.
However, the bonds will be weighted by the agent's definition of
self. In order to be successful, a parochial network must be very
small so that the information structure is efficient \cite{BO2}.
However, in order to play in the entire pool of agents and with
nature, a stratification of skills in the cluster is required if
the cluster is to be self sufficient.  However, the skill depends
on both innate ability and training.  We will be discussing the
idea of aptitude and skill with respect to our model later.
Although training might be imposed on agents by a small parochial
cluster, a wide distribution of innate skill aptitudes is
statistically less and less probable in smaller and smaller
clusters. The other point to be taken into account is that as
agents from different clusters come into contacts, beliefs and
utility curves tend to get modified and the fixed ethnic tag and
the easily flippable beliefs can get less correlated and the tag
can be used to fake beliefs. The other point that might cause the
agents to desert the cluster is mating, which can make the markers
uncorrelated. The choice of mates is not the same as the choice of
business partners where long term commitments are always
necessary.  Also, mating can offer a deserting agent security in
the new cluster so that his self similar variables can get a
guarantee to perpetuate there. Because of all these reasons, we
propose clustering where all variables including beliefs and
utility curves act as possible reasons for bonding, and one agent
can be loosely affiliate with more than one clusters.  The
affinity clusters are dynamic, and hence so are an agent's utility
curves. The other main difference with the parochial model that
takes into account only the ease and cost effectiveness of the
information system of a parochial cluster is that, although there
are bonds that take into account similarity in beliefs that
account for ease in exchanging information, our model also takes
into account similarities in genotypical properties, something
that accounts for unconditional love for children and close ties
with siblings even they grow up physically isolated and share
totally different views.

\subsection{Optimization of Energy}

Each agent in the network tries to attain the maximum utility at
the expense of minimum energy invested.  We can define a variable
similar to free energy to define this term.  The agent will tend
to cause minimum change of entropy of his own environment for the
maximum possible gain of utility.  The cost in this game is
entropy.  We can define this situation by a mathematical
optimization equation with a constraint to minimize entropy.  This
can be achieved if we can find a variable v such that utility is
directly or positively correlated with v whereas entropy is
negatively correlated with v and then we find the maximum of
utility - entropy by varying on v.  Let us call this quantity
(utility - entropy) to be net utility. As each of the agents
maximize their net utility, the social cluster reaches an
equilibrium as no agent can gain more net utility by shifting from
that position.  This situation is similar to reaching a Nash
equilibrium in game theory.   The situation is also similar to a
spin lattice reaching one of its stable phases.

\subsection{Flipping of Variables: Critical net utility and Correlation among
variables}
 The variables can be modeled similar to an array of
spins coupled to one another with weight factors.  These weight
factors can be updated as more information is fed from the
environment and other agents. An agent will tend to maximize the net
utility on each variable by taking the weight factor into account.
When the net utility in one variable is lower than a threshold,
there may be a flip in the preference.  A flip in one of the
variables again affects the other variables depending on the
coupling.  This might be described as follows.  As an agent's
preference shifts, the agent may need to modify his/her skill set to
maximize the new preference.  Also, if the flipped preference
creates a conflict with other preferences, flipping may continue
until a stable state with no contradictions or a minimal
contradiction is achieved.  As a result, the cost of flipping a
variable will be the cost of flipping all the variables that are
strongly coupled to it. The threshold for flipping can be explained
as follows.  In order to flip a variable, a certain amount of energy
needs to be spent.  This energy includes the effect of decoupling
from any affinity cluster that was based on the value of the
variable, energy spent in loss of credibility etc.  Loss of
credibility can be explained as the fluctuations in the interactions
with other spins because of an unstable state of the spin.  Agents
in an affinity cluster will contribute there share of affinity
utilities to the agent only when his variable value is credible.
Once an agent flips his variable and joins another cluster, the
agents of the new cluster will need to calculate whether the flipped
variable is true or bluffed, or whether it is a temporary
fluctuation, since a temporary fluctuation will not make the agent
incur the cost of flipping all the coupled variables.  The other
part of the flipping cost is the physical cost incurred to flip a
variable: the time and energy spent.  Also, flipping a variable
would mean that an agent will be placed in a pool with other agents
with a new similar variable and agents possessing contradictory
variables, if the variable is an axiom, will be acting against it.
Flipping in one variable might also be economically efficient if
flipping in one variable optimizes the utility in some other
variables in the agents array substantially, so that the flipping
cost is expected to be compensated for by future increase in
utilities in other variables. As an agent's variable flips, agents
coupled strongly with the flipped agents on other variables may flip
the certain variable.  This will depend on if the flip weakens the
total coupling of the flip agent from the affinity cluster, and if
the decoupling results in a loss of the affinity utility which is
larger than the flipping threshold. Again, an agent with a flipped
variable may bluff in order not to lose the affinity utility share.
The spin-spin interaction energy among agents can be modeled as
$\sum_i J_{ijk} s_{i}s_{j}$ where $J_{ijk}$ is the coupling constant
between spin $k$ of agent $i$ and that of agent $j$. Flipping the
spin will change the sign of the interaction energy. However, a new
term will be added with the flip, which will be the energy spent in
making the spin flip look credible. Also, the internal energy spent
will be the sum of the flipping energy all flipped spins coupled to
the agent.  If the internal spins are coupled with, weight factors,
then it might be sufficient to flip only leading term spins.  How
many internal spins will be flipped can be decided by the following
equation: We model the flipping energies as follows. The variables
are coupled to similar variables with other agents or the variables
may also be coupled with nature.  If the variables are coupled with
other variables in nature, the total energy in flipping will depend
on the number of agents coupled together in a network, and also the
weight placed on the variable.  Also, the energy required to flip a
variable will depend on the internal couplings.  Let us assume that
the external flipping energies are the product of three variables:

 1. a coupling weight $J_{ij}$, that takes into account the weight
given to the certain variable by agent $i$  $J_{ij}$ is the same
for all $j$ for a specific $i$, as the weight would depend
specifically on agent $i$'s valuation,

2.  the number of agents in the network, or the total number of
nodes in the cluster carrying the same variable,

3. a distance factor that explains whether agent $j$'s variable
have effect on gent $i$'s variable and a constant flipping energy
$E_{flip}$. If, on the other hand, the variables are coupled to
nature, the situation is a little different.  In the cluster with
other agents, all agents are assumed to have the capability of
influencing other agents. However, when a certain variable is
coupled with nature, the flipping energy can be very high. Nature,
here, compared to a large heat bath with which the cluster is held
in contact, and the thermodynamic fluctuations of the clusters
will depend on the spin-spin interactions with the other agents,
whereas, the equilibria in some other variables will be
predetermined by the external environment, determined by nature.
This huge "average" behavior of the large bath will determine  the
unflippable variables or utilities that an agent cannot do away
with.  Examples are eating and shelter.  However, variations on
what to eat and what to use as shelter are not globally fixed, but
depend on local variables and agent to agent interactions, and
these variables can be flipped.

\subsection{The "difficult to flip" variables and mating}

The variables that are difficult to flip can also get reshuffled
from one generation to another by mating.  A network that tends to
optimize efficiency by heavily relying on fixed variables to
deduce inherent values and faith is also prone to get deceived by
"fake markers" carried by progeny of agents belonging to the
network.  In a one generation game where deserters can easily be
tagged by their markers and punished, the fixed markers can be an
efficient choice to enforce homogeneity of values within the
cluster.  However, as the markers diffuse and values get
uncorrelated, the forced belief in the correlation can produce
disastrous results.

Let us imagine a cluster where several sub-clusters marked by some
fixed variables operate within the system.  Let us also assume
that bargaining powers of the agents are based on his/her
placement in the labor market, and also that the labor market is
strictly controlled by the fixed variables, so that a certain job
is done only by an agent carrying a certain variable.  This
situation can put certain sub-clusters to a more advantageous
position as they receive a high bargaining power, and they are
assured of their variables remaining in a high bargaining power
situation over generations if diffusion of markers is rendered
unacceptable.   The situation will create quite a few
inefficiencies:

1.  An agent's aptitude variables may not get propagated in the
same way as the fixed variables.  A guarantee of a placement in
the labor market based on an external marker will also at as a
lack of incentive for individual agents to invest in acquiring the
skill.  As a result, total competence in the skill will go down
for the cluster.

2.  The lack of incentive for the agents placed in lower
bargaining power categories to be efficient.

3.  Dynamic nature of the important skill as the game with nature
and other cluster evolves.  Labor reorganization might be required
to adapt to a dynamic game.  However, ac luster that heavily
replies on fixed variables to assign labor preferences will find
it costly to reorganize quickly, as all systems taking the fixed
variables into account to account for labor must be updated.

4. Two dissimilar variables even kept in the strictest social
rules but in a physically connected space will diffuse. We
elaborate diffusion in a later section.

The other aspect to be taken into account is inter-cluster games,
where because of mating, it might be possible to fake one or more
fixed variables.

The other interesting property of a cluster composed of more than
one sub-clusters held together, even if there is no strict fixed
variable labor stratification is the following:

The subcomponents of the cluster will depend on one another, but
will also need to be held together by a sufficient number of
similar variables that are highly weighted and also have a high
flipping cost.  However, if the sub-clusters differ in one or more
highly weighted variables, they will also compete against one
another when resources become scarce, as each sub-cluster will
prefer perpetuating members of the own clusters that are similar
in more variables.  However, if labor assignments are made within
the entire cluster itself and if all members are allowed to come
into close contacts, the members of different sub-clusters will
have their fixed variables diffuse with generations.  However, any
subgroup possessing a variable that can be initially associated
with a higher bargaining power will be reluctant to give up the
bargaining power unless the other fixed variables from agents from
other sub clusters also hold a bargaining power with an equal
degree.  However, with diffusion, the fixed and skill variables
will get disjoint and unless the variable interpretation system is
totally reorganized, the wrong agent will be affiliated with the
wrong skills.  Now, let us assume that the sub-clusters are held
together for a long enough time without the cohesive variable
being disturbed, and the agents' fixed variables diffuse to create
a cluster where no fixed variables can separate one agent from
another.  What does remain interesting at this part is the
evolution of the set of preferences held by the different
sub-domains.  Each of the subgroups are assumed to hold a set of
common axioms in the beginning that lead to a set of cohesive
decision preferences.   Now each of the sub-domains must have at
least some axioms that are contradictory to the axioms of the
agents in the other sub-domains.  Although genetically fixed
variables can diffuse with no contradictions, diffusion of logical
systems need not be without creating logical systems that contain
contradictory decisions within the logical structure.  Although
initially the contradictory axioms might carry little weight in
the correlated spin structure, these weight factors are subject to
modification with an evolving system placed with other clusters
and also with nature.  As contradictory axioms start carrying
higher weights, the cluster will become less cohesive, and it is
also possible to have weakly correlated genetic markers and
contradictory logic systems or a cluster with simply
self-contradictory logical systems. When the sub-clusters have
unequal labor bargaining power, one sub-cluster might decide to
flip its low cost preference variables in trade of a partnership
and for a logically cohesive cluster.  However, since the axioms
underlying the decision preferences cannot be proved within the
logical system, as long as the diffusing clusters have equal or
close to equal bargaining power, one set of logic tree cannot be
preferably replaced by the other.  Again, an agent willingly
flipping his preference variables will signal a lack in bargaining
power.

We try to simulate a lattice which is initially consisting of
several sub-domains consisting an exclusive set of genetically
fixed variables and an exclusive set of logic systems, but with
equal and different skill bargaining powers.  We let the markers
and the decision preferences diffuse slowly.  The simulation
results and the equation are discussed in the later version of the
paper.

\subsection{Rate of Change of Utility}

The rate of change of utility is used for a corrective feedback
mechanism.  A sudden rate of change in a utility will imply either
a sudden change in flip in others leading to a sudden decrease in
the agent's utility, or a sudden need for reorganization of
variables.  A sudden need for flipping one or more variables
require a large investment in flipping energies.  A sudden change
in utility in a variable will lead to the agent  to adjust weight
factors in other variables while trying to maintain the status
quo.  Also, a sudden change in utility in one variable due to
actions in part of other agents will lead to the acceptance of
defection or betrayal if the agent adapts to the change quickly.
Any cluster must have a built in mechanism to punish defectors. An
agent interprets other agents' weights in preferences by looking
at past data sets of actions, and a data point corresponding to an
act of defection or betrayal will lead to other agents in a
position to gain from leaving the cluster to provide an incentive
to leave.

\subsection{Weight Factors of Variables}

Each agent can be described as an array of an infinite number of
variables.  We consider cases where a finite number of variables
carry a changeable but large portion of the total weights.  On the
other hand, if the weight factors were distributed thinly among
many variables, which are uncorrelated among agents, no clustering
would occur.  However, we consider the weight factors to be also
dynamic. Let us consider $n$ leading variables among $m$ agents.
Let us say that agent $i$ has a weight for variable $k$ to be
$w_{ik}$. In that case agent $i$ will also invest $w_{ik}$ proportion of
its energy in optimizing in variable $k$.  Also, agent $i$ will form
positive or negative bonds with other agents which will contribute
to the agent's "affinity utility" as a function of $w_{ik}$.  However,
if agent $j$ has a weight $w_{jk}$ for variable $k$, agent $k$ will gain
from agent $j$'s affinity utility as a function of $w_{jk}$.  Also,
flipping a variable $k$ will affect an agent as a function of $w_{ik}$
and $\sum_j w_{jk}$ and the alignment of $k$ in other agents.

The weights come into play significantly in the following way:
The agent can only afford a limited amount of change in entropy, or in other words,
the agent has only a certain amount of time and energy available for spending.
This constraint will be taken into account when net utilities are maximized.
The agent will start at optimizing the highest weighted utility, or existence, and will go down the tree
by optimizing utilities that are connected to self by taking the weight factor and the net utility into account.
The variable to be looked at is the utility scaled by the weight factor with the entropy factor subtracted.
When several nodes are reached from one node that represent the same net utility within a certain error range, with the weighted utility put
in, the nodes are pursued in parallel as we go down the utility tree.

\subsection{Weight Factors, Risk Factors and Integration over time}

 We assume that agents placed in a social lattice will play against one another
and also against nature to optimize their utilities.  However,
with every game, we can associate a risk factor.  For example, if
several agents are placed in a market, and each of them values two
different commodities differently, each of the agents will try to
deduce the other agent's valuation in order to maximize his/her
own profits in the futures market.  The other agent's valuation
can be guessed if enough information is collected about the second
agent's past decisions.  However, the utility curve of the second
agent is also subject to change.  As each of the agents interacts
with the environment separately, they acquire more and more
information, and their needs may reflect a changed set of
information possessed by them.  The importance of a certain
utility may also go down or up as new information is added to an
agent's information system.  This possibility of change can be
lumped into a risk factor.

How a certain agent calculates this risk factor also affects
his/her decisions. Again, calculating the risk factor or possible
future actions requires an investment of time and energy.  Since
each agent tends to minimize the energy spent, how much energy an
agent will invest in interpreting the second agent's future
actions will also depend on the first agent's interpretation of
the "importance" of the second agent's actions.

The weight factors are very similar to diversifying ones
portfolio; an estimation of investments made into different
utility-stocks with long term and short term options. Utilities
will be connected with weight factors that will be proportional to
the risk factor associated with the certain utility.  Also,
possible changes are taken into consideration when integrating all
utilities over time.  In a many step game, the expected payoff
from the $n-th$ step depends on the on integrating over all the risk
factors over time.  The weight factor will also depend on the
possibility of the maturity of an $n$ step game.  The other term to
be taken into consideration is the possibility that the utility
variable that the game is optimizing on will not flip by the
maturity of the n-step game, as with a flip, the payoff from the
game will become negative.  Other minor terms to be taken into
consideration are possible inclusions of agents in strong affinity
clusters that will distort the utility curve and shift the value
of the payoff relative to the agent. The existence axiom must be
the most highly weighted variable, which we assume to be fixed.
The existence axiom, as defined in our previous paper, can be
described in detail as an agent's utility in perpetuating the
array of weighted variables in the closest possible unchanged form
so that the highest possible utilities are obtained from the
variable, taking the weights into account.  However, when the
utility fall below a threshold, a variable can be updated, as the
utility is not contributing to the existence axiom then.  If,
taken the flipping energies into account, a flipped variable
produces a higher than threshold utility, the flipped variable
will redefine the definition of self. Again, some variables are
connected to the self axiom with a high flipping energy threshold
and also a difficult to modify large weight factor.  These are the
variables that connect the agents with the environment or nature
in a material way, so that flipping them will inevitably cease the
existence axiom.  For example, eating or shelter are utilities
that are very difficult to flip, though the preference in eating
might be somewhat modifiable.  So some variables that are used for
linking with nature have a high flipping energy.  This is somewhat
similar to a spin system being linked to a larger thermodynamics
system where the variables are controlled more by the larger
system's average than the individual fluctuations of a small
system connected to it.  The utility variables linked with the
existence are again optimized because they are expected to
perpetuate the existence. The existence axiom will also take into
account the coupling strengths among the variables when defining
the meaning of existence.  For example, the variables inside an
agent are inter-connected closely, and flipping one of them
effects other variables in the agent's array strongly.  However,
the couplings with other agents' variables are long range, and an
internal shift in an agent's variable will have a long range
effect in other agents' similar variables.  As a result, the idea
of self is concentrated most within the physical agent himself and
fades away as longer and longer range couplings, and also
couplings with agents with more and more dissimilar variables are
reached. Some of the weights might be easily shifted, whereas
others might have a hard shifting possibility.  For example,
material needs such as food and shelter have a high weight factor
determined by nature.  These weight factors depend on the agent's
game with nature rather than the agent's interaction with other
agents.

\subsection{ Mutations}

 In this term, we try to explain the
possibility of the mutation of a utility curve or preference.  The
mutation term explains sudden changes in utility curves in random
agents.  This sudden change occurs in individual agents due to
local interaction with nature or other external factors instead of
interactions with other agents.  Mutation, hence will be a flipped
spin or a new axiom connected to the tree of an individual agent
by local interactions with nature, and may occur even when the
thresholds to flip have not been reached. Let us focus on the
difference between the regular flip by reaching a threshold low
utility in the flip variable and a mutated flip.  A mutated flip
will occur regardless of the current utility in that specific
variable.  As a result, two things can happen:

1. The process may simply speed up a change that was slowly taking
place, or a change in utility which was happening very slowly at
the cost of efficiency to the whole system.  For example, if a
system is held right above the threshold utility in a certain
variable for an indefinitely long time, naturally, there will be
no flip in the variable.   However, the integration over a long
time in maintaining the variable at a value just above the
threshold may be in general more costly to the system in total
energy minimization than a mutated flip that would reorganize the
system. We can write down the equation as

2.  A mutated flip in a variable that is already optimized will
cost a flip in other variables or cause a huge contradiction.

Later the change may or may not diffuse across the network,
depending on the parameters in the diffusion network and the
specific advantage against nature and other clusters gained from
adapting to the new axiom and the cost of replacing older axioms.
As a result, the probability of a cluster undergoing sudden change
$=P(mutation)+P(diffusion)$.  The diffusion of any mutation will
be opposed by any agent with a high investment in a variable
contrary to the mutated variable. Now, the probability of the
diffusion across the cluster to other clusters will depend on the
relative gain of the mutated cluster against other clusters by
holding the mutated variable to themselves. When a mutated flip is
actually reducing the total efficiency of the system, a diffusion
will be resisted by most of the agents.  Even if the flip is
increasing the efficiency of the system, diffusion will be
resisted by agents who have long term investments in the unchanged
variable. The other type of mutation is the addition of a new
axiom or preference.  This new preference may or may not be
contradictory to the existing preferences.

\section{RISK AND AXIOMS}

\subsection{Risk, Insecurity and Spurious Axioms}

As the agents interact with nature to find more axioms that become
fixed with nature, more axioms remain to be found. Also, there is
no such determined linear correlation among knowledge,
calculations and the actual gains.  In every decision, some risk
factors are associated, and some of these risk factors pose minute chances
of gain against huge odds.  If we examine history, most social
faith systems were created on the verge of deaths and extreme
decays of societies where large risks were required with odd
gains.  An agent can invent his own faith system in order to
create a virtual gain that is guaranteed if a certain risky action
is carried out.

If we look at possible futures in the point of view of the many
universe theory \cite{EV1}, and possible decisions with high and low risk
factors, with unknown or unexpected results, a rational human
being will always choose the decision with the highest expected
utility which will usually be associated with low risk if the risk
factor contains a non diversifiable portfolio of investment such
as the cost of an agent's life.

However, this choice of decisions may not be the best possible
choice for a system as constant low risk decisions must be
associated with constant slow changes.  However, a high rate of
change in utility may not be countered by slow changes.  As a
result, high risk decisions may be required to move a system from
a fast decaying utility curve to a stabilization by adding risky
pieces of knowledge.

\subsection{Scarcity and Non Fixed Preferences}

 When an agent fails to meet the minimum threshold in an unflappable axiom,
 he first tries rearranging the weights so as to minimize the effect of the loss.
  However, if the weight of the certain axiom is also fixed, then as a strategy,
  the agent might create a spurious set of axioms to add to the
  "existence" axiom so as to keep it from flipping. The spurious set
  of axioms might be contradictory to the original set of axioms,
  and might, with time, fail to correct for a situation where possibility
  of optimizing the original utilities has been restored. Just as any
  other set of axioms, the change of a spurious set of axioms will be
  opposed by the agent's tendency to maintain the status quo.

\subsection{Diffused Logic Systems and Inconsistency}

A diffused logic system will contain diffused axioms from both
pure logic systems.   However, a diffusion from both parts will
occur with a low resistance only when high weight terms do not
contain contradictions, or forcibly one set of logic terms are
chosen over the other to maintain consistency.  However, any well
developed logic systems containing many preferences based on its
axioms must allow diffusion of contradictory terms in low weight
positions as sorting and correcting inconsistencies in all terms
will take a huge amount of time and processing power.  These new
logical systems may then be passed from one generation to another
as a given faith system or a system of axioms that are inherited
or taken for granted.  Over generations, the visible tags of the
two populations can get mixed to produce a homogenous population
with an almost homogenous logic system with contradiction only in
minor low-weight terms.

The interesting phenomena occurs as with time and interaction with
nature and other clusters, the weights for the preference terms
need to be modified.  As the spin array with its associated
weights is allowed to evolve, we may come across phenomena where
the low weight terms ignored for inconsistency correction purposed
during diffusion become leading terms.  This may happen due to a
sudden scarcity, or a new knowledge acquired from nature.

If the weight factor is changed with leading inconsistencies in a
person's logic system or in the logic system of agents placed in
the same logic pool, as the contradicting axioms will tend to make
macroscopic changes with costs in entropy but leading to opposing
macroscopical changes, conflicts must arise.  The magnitude of the
conflict will depend on the agents' perception of the weight of
the contradictory preferences.

\subsection{Creation of New Logic Systems and New Faith}

A faith system can be created by choosing a set of axioms
exclusive to the axioms already possessed by an agent, given that
the faith set of axioms do not contradict with the existing axioms
or preferences.  A faith system may carry a very high weight
depending on whether it is connected to an axiom with high
flipping energy.

As an agent interacts with nature and other agents to acquire new
"collapsed" axioms shared with other agents, the faith system
always has a chance of possessing contradictions to one of the
newly acquired collapsed axioms.  However, since the collapsed
axioms are shared by agents in the clusters, and may be coupled to
nature with very high flipping energies, contradictions with an
existing faith system may create large contradictions within the
agent's logic system.  If the faith system is coupled with some
other variables with high flipping energy, but not shared by all
members of the cluster, then the contest between the faith axioms
and the newly acquired axioms will be subject to cluster efforts,
as sharing a common axiom will lead to strong clustering and group
efforts in maximizing utility in that certain preference axiom.
Now a faith axiom can be discarded by an agent who is able to
devise a new set of faith axioms that is not contradictory to the
agents redefined set of axioms.  However, the shift in the faith
axioms must also be justified by another set of axioms that do not
tend to flip the vital axioms the faith axioms are coupled to.  In
a nutshell, a substitution of a faith axiom must be designed so as
not to disturb the agent's value of the vital spins.

As more and more collapsed axioms are acquired, entropy in nature
increases.  However, this leads to a more complicated system.  If
there are only finitely many axioms to be acquired from nature,
then after an n step game, all the rules of the game would be
acquired and knowledge would be complete.  This situation will
imply maximum entropy in nature, as absolute knowledge would imply
absolute knowledge of the future, and also no choice in future
moves.  In any such system, no spurious system of axioms can
exist, as all axiom or anti axiom will already be acquired, and
hence there can be no faith system.  However,

\section{Defining the Entropy Factor}

 The entropy factor takes into account the disorder created in the agent's environment.
Now this entropy may not be a simple function for disjoint states.  To start with,
this entropy must takes into account long range correlations among
matter.  For example, an increase in entropy at one space point
may appear as an increase in entropy at another point because of
the connectedness of events.  In a simple system where long range
correlations are present, often Tsallis entropy can take into
factor the corrections corresponding to the correlations. However,
an entropy associated with a more sophisticated network where each
of the agents are themselves complex agents should intuitively be
more complicated. In a specific way, the entropy is the agent's
calculation in the damage caused by either disturbing the
ecosystem.

In the mathematical term, the first utility term could also be
called entropy, as an increased utility is what an agent perceives
as negative entropy. So an agent tries to maximize negative
entropy and minimize entropy. Any action that decreases an agent's
utility is actually causing an increased entropy. However, we have
separated the two terms for the following reasons:

1. utility may include terms that are completely non materialistic
and clauses like beliefs and faiths. These terms may simply
reflect an agent's estimation of insecurity and risk, and not the
proper entropy.

2. the utility terms are weighted, and may put a very low weight
to a term that is causing a high entropy change in the
environment. However, optimizing in one of the utility terms may
come at the cost of suddenly or slowly lowering other terms. Most
agents would put a higher weight on immediate utilities than on
long term utilities because long term utilities have higher risk
factors associated with them. However, small constant changes in
the low weight entropy  factor may at one point exceed a threshold
that causes other utility functions to flip because a critical
increase in entropy interferes with the optimization of one or
more vital utility factors.

So we take entropy to be a switch like function connected to one
or more vital utilities. The model behaves as follows: The entropy
function usually increases slowly with any action the agent
carries out.  As a result, the entropy term is very low compared
with the agent's other cost functions such as energy and time
spent.  So entropy does not play a role in an agent's utility
weights.  However, as the entropy function is connected to several
vital utility functions, so that when the entropy function reaches
a certain threshold, one or more utilities with very high flipping
cost, when total or accumulated entropy reaches a warning level,
some of the less vital utilities are flipped to minimize the
change in entropy in the local environment.  The other two things
to be taken into account here are the local and global nature of
entropy and also the time steps. Entropy can be split into two
parts, local and global.  By optimizing the utilities connected to
the existence axiom, an agent tries to persist longer, or decrease
local entropy.  In order to do that, he must interact with nature
or gain knowledge from nature.  By doing so he increases global
entropy. The decrease in local entropy must be a constant process,
as the game is played against nature, which is increased in
entropy continually, and a pause in the action will increase local
entropy.  However, an action carried by an agent alone will cause
a fractional change in global entropy and will thus have a low
effect on the agent's own local entropy.  If rate of change of
global entropy as opposed to the rate of change of local entropy
due to an agent's action is lower, the global entropy will have a
low weight in the agent's action, ie. An agent will do nothing to
offset the cost of the increase in global entropy caused by his
action. However, small constant changes in the low weight entropy
factor may at one point exceed a threshold that causes other
utility functions to flip because a critical increase in entropy
interferes with the optimization of one or more vital utility
factors.

So we take entropy to be a switch like function connected to one
or more vital utilities that have a high cost function for
switching; hence, exceeding a certain value in the entropy will
require one or some of the high cost utility functions to flip,
making the agent incur a large cost.  This can be balanced by
flipping several other utility spins.  The spins will be chosen
such that the immediate entropy increase can be minimized at the
minimum flipping cost.  The time factor is taken into account here
because of the risk factor associated with the maturity in any
long term investment.

\subsection{Short vs. Long-term Risk in Local and Global Entropy}

Minimization of local entropy is in many cases connected with
short-term gains.  However,  increase in global entropy is a slow
process and is associated with long term risks.  Hence, the energy
invested in the minimization of local entropy vs that in
calculating the cost of global entropy is that of investing in
short and long term stocks.  A long term investment may or may not
mature (an agent may die), or a long term risk may be countered by
other branchings in actions and technology.  The longer the
investment is for, the more the chances are of the result being
affected by probabilistic changes.

\subsection{Imposed Check in Global Entropy}

A check in global entropy can be forced by associating with the
local entropy function.  However, any such association will come
with the cost of the allocating an agent's limited time and
resources to calculating global entropy at the cost of optimizing
other local utility curves.

\section{SPECIFIC EXAMPLES CONCERNING CLUSTERS}

\subsection{Clustering and Sub-domains in a Stable Phase}
 If we consider the labor market, and the division among jobs, several
factors need to be taken into account:

1. An agent's aptitude for a job, which again depends on
   a. The agent's propensity for that skill
   b.  Investment made in acquiring the skill

2. The other agents' aptitude in measuring the skill

3. The demand for the skill This process can be summarized again
as maximizing the net utility of each of the agents.  However,
there are several factors that are noteworthy here:

1. Net utility is not the same as utility.  When an agent is
trying to find the maximally skilled agent for a job, he also
needs to invest energy and time for the search. An agent will try
to find the most skilled for the minimal energy spent.  If an
agent calculates a generalized correlation among two variables,
one of which is hard to measure and the other easy, as long as the
cost for neglecting exceptions are calculated to be not very high,
an agent might tend to measure the easily measurable variable to
deduce the value of the more difficult to measure variable.
However, since each agent also is trying to maximize net utility,
one agent's inaccurate measurement of a skill might reduce the
total net utility of other agents who have a demand for that skill
and will share the skill.

2. An agent's skill depends on an agent's affinity for a skill and
also the agent's investment in acquiring more information to
improve the skill.  Again, an agent with an affinity for a skill
will need to invest less to acquire mastery in that certain skill.

Creation of Labor Clusters and Tags: security and Long Term Risks
The resultant clusters in the labor market might not necessarily
reflect every individual with the optimal skill at the most
appropriate job sector because of each individual's tendency to
maximize net utility and not utility itself.  Also, besides
utility, we must include a security term in entropy. The problem
of security is also very closely related to entropy and the
"existence" axiom that we argued about in the previous paper. A
labor domain is not necessarily the same as an affinity domain.
Again, a certain skill is only one of the many variables that can
be coupled into a group.  However, many variables have no
correlation with respect to games with nature or "quantitative"
utility whereas some others do.  As a result, affinity domains and
labor domains will be two different coverings of the set of all
agents. In order to play successfully against nature and
environment, it is efficient to create forced clustering in
skills, or trade skills.  However, the agent whose skill is
important may or may not belong to a strong affinity cluster with
the other agents in the trade.  Hence, the clustering here is
"forced" on need, with obvious possibilities of betrayals in the
last step of any game.

The following equations can succinctly describe the game: An agent
will make an investment of $I = f(A) + g(tagging) + h(demand)$
where $f(A)$ is a function of aptitude.  A person with higher
aptitude will need to invest a smaller amount in order to achieve
the same skill level as a person with lower aptitude.  A person
with lower aptitude may not be able to overcome a threshold in
acquiring the skill. $g(tagging)$ is based on the tagging barrier.
An agent carrying a variable which is difficult to flip, but
acting as a tag against a certain variable will need to invest an
extra energy equivalent to the tagging  potential in order to be
credible as a carrier of the certain skill $h(demand)$ is simply a
function of demand.  A skill with higher demand will yield a high
pay.  Hence investing in a skill with a high pay will yield a
higher degree of freedom and more free time that can be used
elsewhere. An agent in charge of assigning a correct labor
position to the qualified agent will try to maximize gross
efficiency in terms of the minimum effort spent.  The utility of
the assigner will be a function of the aptitude of the candidate
and the difficulty in interpreting the aptitude.  This can be
written as $U(Assigner) = f_1(u,tag)$.  $f_1$ is a function that
is dependent both on the total aptitude of the candidate and the
difficulty in finding the aptitude. The total utility of a person
only benefiting from the labor of an agent is
$U(consumer)=f_2(u)-f_3(cost)$ Where $U$ is only a function of the
assignments of the assigned agents.  As a result a consumer will
be willing to invest in a cost term that will punish the assigner
against a gross tagging scheme that will bring down the total
labor skill, as long as the total salary paid for the skills
purchased is the same. The total utility of a competitor will be a
function of both his aptitude and the negative of tagging against
his competitor. However, when we look at the utilities of the
assigner and the consumer, we see that tagging must go down as the
aptitude of the candidates fall far outside median mainly because
of two reasons:  (a)  candidates with aptitudes far outside median
will be easily discernible, (b) eliminating candidates far outside
the median will bring down the total labor to the consumers by
larger amounts, and hence excluding candidates with high aptitudes
by tagging will be automatically checked by consumers.

However, the correction terms imposed by the effects of clustering
have yet not been imposed. A further game can be developed if we
connect negotiating power of the agents with these variables.  A
detailed game will be sketched in a paper being developed.

\subsection{Forced Clustering}

Let us assume that cluster $A$  and cluster $B$ have two
dissimilar variables in location $x$, and that clustering is
created by filtering on that variable.  However, let us also
assume that in both the clusters, there are two other variables to
be considered: an easily discerning variable, $y$ and a skill
variable $z$. Now let us assume that there is a weak correlation
between $z$ and $y$, so that agents with $y$ are correlated with
$z$ and is discriminated against when skill $z$ is considered,
however, we also consider that there is no correlation between $x$
and $z$. Now any skill is a combination of both aptitude and
investment in the skill.  If cluster $A$ is in a more economically
advantageous position, so that cluster $A$ can pay higher for
skill $z$, then cluster $A$ can cluster agents in the basis of $z$
in cluster $B$ and offer higher pay to the $z$ agents possessing
$y$ as well.  Then a forced clustering on the basis of $y$ with
the expectation of high affinity contribution with respect to $y$
will serve the purpose of cluster $A$, as long as $y's$ in cluster
$A$ have a higher affinity for variable $x$ than variable $y$. Now
agents from different clusters or domains are discriminated
against over the possession of a certain variable does not mean
they can naturally be forced into a viable cluster. Agents who are
expected to pay dearly because of the forced clustering at the
cost of agents unskilled $z$ might as well prefer getting isolated
and sell their independent skill in $z$ in a strict one step
negotiation. The other very important part here is whether a weak
correlation with the lack of skill $z$ in agents possessing $y$
implies the existence of say a weak correlation with skill $w$ and
a natural preference and inclination towards that skill. In a
later paper we expand in the possibility of valuations of skills
and negotiation power and design a game based on forced
clusterings of specific variables. Also, since the $y$ tag biased
discrimination is applied to the initial evaluation process, where
the evaluator must apply extra energy in identifying the few
qualified $y's$, if the qualified $y's$ in cluster $B$ get
isolated and by a higher investment than average make her skills
credible, or have an aptitude above a certain threshold, where the
$z$ aptitude is clear, and requires little investment in
identifying the aptitude, then it might be more advantageous for
an agent to not join the join the forced cluster. In these
situations, it might be efficient for agents without $y$ to
include the minority distinct isolated agents. This assimilation
will also depend on the long term effect of the perpetuation of
the $y$ tag.  If the probability of an agents' perpetuating the
$y$ tag does not depend on the agent possessing it, then there is
no long term threat in an agents' self perpetuation by including
the exceptional $y's$.  However, if tag $x$ has a correlation with
the agent possessing it, then a clustering in $x$ will have a more
long term effect that will span generations than a clustering in
$y$ for skill $z$, which will be merely a function of efficiency
in the investment made in isolating the skill.  Now the forcible
clustering in tag $y$, when a clustering in $x$ already exists
will prove to be advantageous to the agents in cluster $A$ as long
as there is no guarantee that the variables with the highest
weight factors of the incoming $y's$ from cluster $B$ are will be
perpetuated in cluster $A$.  Again, any such assimilation of
variables comes with the usual risks of defection. When the
defection is associated with the defection of easily discernible
variables, at the cost of flipping the easily flappable variables
to offset the total price, the long term risk remains the
probability of the entire clusters undergoing phase transitions in
the easily flappable variables, keeping the difficult to flip
variables as the dissimilar variables, and the possibility of that
dissimilar variable passing from generation to generation.

We try to model the situation with the following equations Let us
assume that $f(x)>f(y)$ By $f$ we mean the filtering effect. This
will happen when

1. $J_{ijx} > j_{ijy}$
  and
2. $\sum Sl_{xl} > Sl_{yl}$

   (here $S$ is the internal spin weights between nodes $l$ and $x$ or $y$)
for majority of agents.

Now let us assume that a clustering in variable $x$ distorts the
preference curve of $y$ by $\epsilon_{I}$ for agent $I$ and a
clustering in $y$ distorts the utility curve of agent $I$ wrt $x$ by
$\Delta_{I}$. Now if $\epsilon_{I}$ $<$ $\Delta_{I}$, then a forced
clustering w.r.t. $y$ will create sub-domains within the cluster
with strong affinities outside the $y$ based domain.

Now let us assume that since $y$ agents are tagged against $z$,
only the exceptional members of $y$ are seen in $z$, and the
number is equally distributed among both $x$ or $y$, or since the
tail part of a gaussian distribution of aptitude is sparsely
distributed, let us assume that there are only few $y's$ in $z$ at
a certain time point which might have equal or unequal
distribution of $x's$ since the sample space is small.

Now let us assume that some $y's$ are clustered together based on
the tag, and are forced to contribute equally to the cluster. They
are also given a quota or a proportion of the total $z$ labor
market.

For the exceptional $y's$: personal tagging cost = 0 or very low.
As a result, the equations for the exceptional $y's$ are cost for
imposed distorted utility contribution in forced cluster $C$, cost
for credibility of actual aptitude in a forced cluster with large
shared utilities $= C_C$, cost for extra competition in a cluster
where most agents would have been cut off from competition in the
absence of the forced clustering $= C_{co}$, cost  for competing
against affinity clusters of competing sub-domains to force one of
their affinity agents into a predetermined quota based position $=
C_A$, gain from being able to gain a position with lower
investment in acquiring skill $= G_L$ cost for being in the same
category as agents who has acquired the same low skill level, and
are backed by their affinity clusters $= C_L$ cost for joining an
affinity cluster based on $x$ in order to compete with lower
skilled $y's$ from the competing affinity clusters $= C_{Cl}$.

An agent with exceptional aptitude will agree to the forced
clustering situation only when all the costs of joining the
cluster are offset by the gain. If the $y$ tagging is not
guaranteed to propagate to his/her progeny with a higher
probability than those with no $y$ tagging, there is no long-term
gain from joining the forced cluster.

\subsection{An Affinity Cluster in Peril in Terms of the Spin Model}

An affinity cluster may for many reasons be at the point of
dissolving.  In his recent paper \cite{BO1}, Samuel Bowles  tries
to explain why altruism in some agents is an evolutionary stable
strategy by arguing about how when at the verge of dissolving, it
is economically most advantageous to leave a society, it is
altruism that can keep a society together.  Again, in his book,
"The selfish gene" Richard Dawkins \cite{DA1} argues that altruism
is an imposed behavior which actually reflects the self interest
of the "selfish people" of the society.  Here, we try to blend in
the modified definition of self and the affinity utility to these
ideas. (how do agents react at that point.. when will they leave.)
Let us imagine an agent $A$ in a cluster $C_A$ at the point of
dissolving.  As we argued before, as long as the agent was not a
bluffing agent, and the cluster boundaries were not drawn
forcibly, the agent will contain a reasonable amount of weighted
variables in common with the other agents: either genotypical ,
philosophical or a mixture of both. Now, let us imagine that the
cluster is at the point of breakdown. i.e, let us say that most of
the agents attain a negative net utility from being in the
cluster.  In classical economics, one would argue that it is
economical for the agents to leave the cluster at that point.
However, here, we take into account a few more correction terms.
First of all, to be accepted as a member of another cluster, an
agent must have or at least credibly bluff to have some variables
in common, as the "affinity utility" of the other agents used by
the new agent must be offset an incoming affinity utility.
Otherwise, the common utility shared by the agents in the cluster
needs to be balanced by an incoming trade or gain.  This incoming
gain in the lack of an affinity term will come with the risk of
betrayal at any step and leaving the cluster.  If, on the other
hand, and agent flips some variables, each flip comes with a cost.
Now the agent can leave the cluster and join another cluster only
when the total cost of flipping and making the flips look credible
to the incoming cluster is affordable.  Then again, a flip changes
the credibility of the variables as the incoming cluster gains the
data about at what cost the incoming agent is willing to flip
variables and move away.  It also comes with the cost of the
outgoing cluster losing the affinity term from the deserter, and
the cost of the "trust" factor that allowed the agent to share
other agents' affinity points.  A deserting agent signals either a
flipping or a bluffing. However, altruism is defined as an act
that is done by an agent in order to help optimize the utility
curves of other agents with no expected return.  Even, when we
take the cost of flipping the variables and the risk of losing
credibility into account, the total investment made in part of the
altruistic agent may actually overweight the hidden cost incurred
by simply not being altruistic.  For example, an agent may simply
sacrifice his life.  There is no further step in the game for the
agent himself, and hence so no expected return.  The existence of
at least a few such altruistic persons has been proved to be an
evolutionary stable strategy \cite{BO1}.  This phenomenon can be
described by in two ways at least in terms of our spin model 1. In
our previous paper, we described the existence of spurious axioms.
Spurious axioms may be traded or even generated by the agent
himself in order to compensate explain any insecurity connected
with the existence axiom \cite{FS1}.  A spurious axiom might
contain a clause that might make the agent inclined to be
altruistic.  A spurious axiom will be successful in over-weighting
an agent's self perpetuation axiom only when it can propose
another way of self perpetuation that is stronger and even after
weighing over the risk factors provides a higher chance of
perpetuation for longer.  Hence, in the agent's own logical
system, an agent will act completely rationally to optimize his
"self perpetuation." However, since all agents do not share the
same set of spurious axioms, the act can appear as irrational to
other agents.

2.  The distorted idea of self can account for altruism.  Many
models involving genes and memes have been proposed in this regard
\cite{HE1, DH1}.  We propose a similar but slightly modified
version in terms of the spin glass model.  An agent is described
as an array of variables.  However, unlike previous models, out
variables in the definition of self are weighted, and the weight
factors can be modified.  The variables themselves can be flipped.
Some variables are inherited genetically and have a high flipping
energy, whereas some other variables are axioms or utilities and
faith.  These can be flipped at costs dependent on the
interconnection of these variables with other variables, and also
the cost of credibility.  A person will have a distorted idea of
self, that will take into account his own array of variables
modified by a weighted factor of other agents possessing similar
variables.  An agent will try to perpetuate his own array, and
will modify the values of the variables in his array according to
the costs incurred in order to protect the entire array.  However,
at certain steps the necessary modifications needed to flip the
number of variables in order to retain the integrity of the entire
array may be higher than the amount of energy and time available.
At that point, the agent might choose to invest in perpetuating
similar variables in other agents at the cost of his own array.

\section{Simulation of a Toy Social Network}
Let us now consider a cluster of agents $\{a\}: a=1,2,...,N_a$, each
with the attributes $\{i\}: i= 1,2,...,N_s$.

Let us consider as the negative of the utility the function
\begin{equation}
H = (1/2)\sum_{aij} A_{aij} s^a_i  s^a_j + \sum_{abij} B_{abij} s^a_i
s^b_j +  h \sum_{ai} C_{ai} s^a_i
\end{equation}
Here the superscripts $a,b$ indicate the different agents,
and the subscripts $i,j$ denote the particular attribute and the
state is a "spin" state which we keep within $\pm s_{max}$, to
indicate the projection along a natural axis.

The $A$ term indicates the coupling of the different attributes in
a particular agent, the $B$ term stands for social interaction
between different agents affecting the attributes of each agent,
and the $C$ term used to represent the interaction of with nature $h$
of each agent with a particular set of attributes.

Taking the derivative with respect to $s^a_i$ of H indicates how a
change of the attribute spin towards a more positive value will
change the utility, i.e. in our simulation we shall increase $s$ by
a unit if the derivative is negative, and diminish it by unity if
the derivative is positive:
\begin{equation}
\partial H / \partial s^a_i = \sum_{j} A_{aij}s^a_j + \sum_{bj} B_{abij}
s^b_j +  h C_{ai}
\end{equation}
determines the augmentation of the spin upto $s_{max}$ or its
decrease upto $-s_{max}$. In our case we experimented with the
$A$, $B$ and $C$ matrices, after taking the simplest nontrivial
set with $N_a = N_s = 3$. We also added a random component to the
matrices as is appropriate in a spin glass (\cite{SK1,EA1}, and as one might
expect in a social context from inexplicable sources.
\begin{figure}[ht!]
\includegraphics[width=8cm]{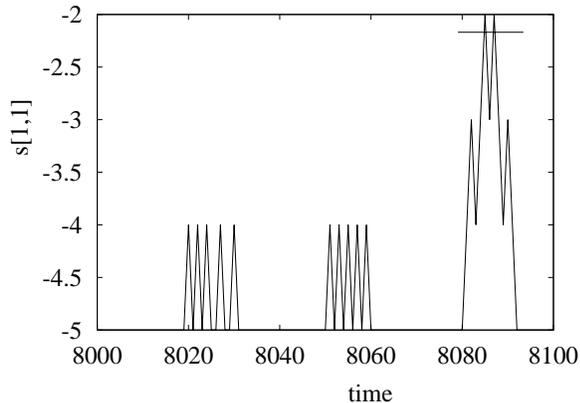}
\caption{\label{fig1}Orientation of attribute $1$ of agent $1$ as
measured by its projection along a natural axis}
\end{figure}

In Figs. 1-3 we show the time evolution of the attribute {\em i}
of agent $a$, i.e. of  $s^a_i$ for three different combinations of
$a$ and $i$. The other 6 possible figures are quite similar to
these three for our choice of the link matrices.

We note the interesting fact that in Fig. 1 for $s^a_1$ we see
static phases alternating with rapid fluctuations, which reminds
one of punctuated equilibria (\cite{RA1, GU1}). However, the analogy is not exact,
because in our case the static states correspond to a constrained
extremum value of the spin or the first agent's first attribute's
orientation in natural space.
\begin{figure}[ht!]
\includegraphics[width=8cm]{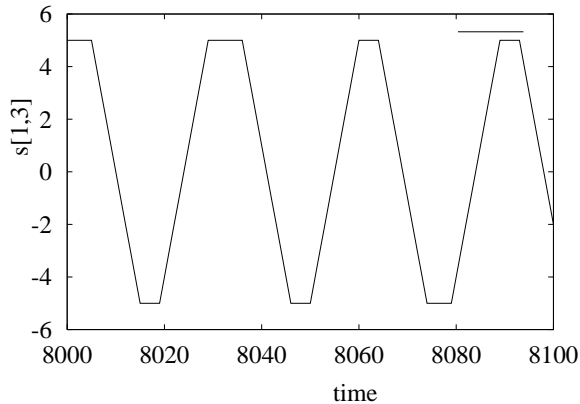}
\caption{\label{fig2}same as Fig. 1, but for attribute  $3$.}
\end{figure}

In Fig. 2 we see a virtually periodic oscillation with no sign of
the small random external noise fed into the different
interactions. This corresponds to the third attribute of the same
agent, whose $A$ and $B$ matrix elements are now taken to be different
from those for attribute $1$.
\begin{figure}[ht!]
\includegraphics[width=8cm]{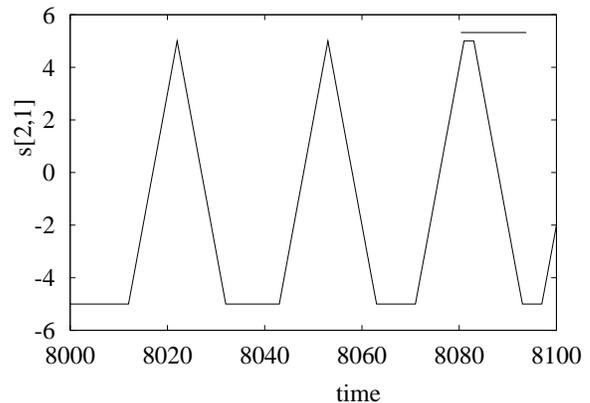}
\caption{\label{fig3}projection of agent 2's attribute 1}
\end{figure}

In Fig. 3 we show $s^2_1$, i.e. the first attribute of the second
agent. Here too the orientation shows a periodic oscillation. But
the pattern is somewhat different from Fig. 2, because there we
have short static periods with alternate orientations, whereas in
Fig.3 we have quasi-static orientations only in a particular
direction, the oscillation to the other direction being relatively
short-lived.

\section{Conclusions}
In this report we have outlined the possibility of
developing concepts and relations related to the evolution of
social clusters, analogous to spin systems, and the importance of
the concept of the "self" of each agent with variable attributes which may be
quantifiable. Simulations with weight factors for different couplings between
agents and their attributes and spin-type flips in either direction from
consideration of a utility function in a simple toy system seem to show the
possibility of  chaos, or at least highly aperiodic behavior, with
also the possibility of punctuated equilibrium-like phenomena. It
would be interesting if the reverse process of obtaining the $A$ and
$B$ matrices from real data can be successfully realized. However,
because of the very large number of parameters available, it would
probably almost always be necessary to reduce the problem to
simpler systems with a manageable set of matrices of links, using
assumptions of fuzziness or symmetry or some other consideration.

\begin{acknowledgements}
I would like to thank Professor P.W. Anderson at Princeton and
Professor Samuel Bowles of Santa Fe and Professor Josh Angrist of
MIT, as well as Professor Bertrand Roehner of the University of
Paris for useful discussions.

\end{acknowledgements}

\appendix
\section{Overlapping Perception and Utilities }

 Einstein spoke of an ant going on in a circle on the surface of an orange and
 assuming that it is traveling on an infinitely long one dimensional
 line.  The other example he gave was of a two dimensional being
 living on the two dimensional shadow of a three dimensional object.
 The knowledge of the two dimensional being will be bound by
 its perception of the projection of the three dimensional world
 onto that two dimensional surface.  Our perceptions are also bound by what
 we can feel from the world, and our logical system is constrained by the
 axioms or information obtained by these perceptions.  Now, the fact
 that most human beings can communicate in a rational manner is
 supported by the idea that a large portion of these perceptions
 are overlapping.  However, we cannot prove within a system bound
 by our perception that some portion of our perception is different from the perception of
 other beings.
 We cannot disprove that the projections of a higher dimensional world that we
 perceive are exactly the same for all agents, if we are a 3-1 dimensional world embedded in
 a higher dimension.
 The difference in human expectations and fine understanding lead to the idea that we are
 able  to project slightly varied images of a world which might be embedded
 in a higher dimensional manifold.


\begin{thebibliography}{}
\bibitem{FS1}F. ~Shafee, "A Spin Glass Model of Human Logic Systems",
{\it http://xxx.lanl.gov   nlin.AO/0211013}  (2002)
\bibitem{GO1}B. M. T. K. ~G\"odel, {\it On formally undecidable propositions of
 Principia Mathematica and Related Systems}, (Dover Pub., reprint
 edition,1992)
\bibitem{VO1} J. ~von Neumann and O. Morgenstern, {\it Theory of Games
and Economic Behavior}, (Princeton University Press, Princeton,
1944)
\bibitem{BO1} S. ~Bowles and H. Gintis, "Persistent Parochialism:
Trust and Exclusion in Ethnic Networks", forthcoming in {\it
Journal of Economic Behavior and Organization } ( 2003)
\bibitem{BO2} S. ~Bowles and H. Gintis, "The Evolution of Strong Reciprocity:
Cooperation in Heterogeneous Populations", forthcoming in {\it
Theoretical Population Biology} (2003)
\bibitem{LM1} P.F. ~Lazarsfeld and R.K. Merton,
"Friendship as a social process: A substantive and methodological
analysis", in {\it Freedom and Control in Modern Society} ed. M.
Beger et al. (Van Nostrand, 1954) p.18
\bibitem{DA1}Richard ~Dawkins,{\it The Selfish Gene}, (Oxford University Press, 1989)
\bibitem{HE1} K. ~Henson, "Memetics",  {\it Whole Earth Review }  {\bf 57},
50 (1987)
\bibitem{DH1} D. ~Hofstadter, Metamagical Themas (Basic
Books, 1996)
\bibitem{EV1}S. ~DeWitt, R. Neill Graham, eds,  {\it The Many-Worlds Interpretation
of Quantum Mechanics:}  Bryce Princeton Series in Physics,
(Princeton University Press,  1973)
\bibitem{CO1}J. D. ~Collier, "Information Originates in Symmetry
Breaking", {\it Culture and Science}   {\bf 7}, 247 (1996)
\bibitem{SK1} D. ~Sherrington and S.K. Kirkpatrick, {\it Phys. Rev. Lett.} {\bf 35}, 1792 (1975)
\bibitem{EA1}S.F. ~Edwards and P.W. Anderson, {\it J.Phys.} {\bf F5}, 965 (1975)
\bibitem{RA1} D.M. ~Raup, {\it Science} {\bf 231}, 1529 (1986)
\bibitem{GU1} S.J. Gould, {\it Nature} {\bf 366}, 223 (1993)
\end{thebibliography}
\end{document}